\documentclass[sigconf,natbib=false]{acmart}

\usepackage{amsmath,amsthm,mathtools}
\usepackage[english]{babel}
\usepackage[T1]{fontenc}
\usepackage{csquotes}
\usepackage{graphicx}
\usepackage{subcaption}
\usepackage[datamodel=acmdatamodel, style=acmnumeric]{biblatex}
\usepackage{xcolor}
\usepackage{nicefrac}
\usepackage{float}
\usepackage{balance}

\AtBeginDocument{%
  }

\copyrightyear{2023}
\acmYear{2023}
\setcopyright{none}
\acmConference[WPES '23 (to appear)]{Proceedings of the 21st Workshop on Privacy in the Electronic Society}{November 26, 2023}{Copenhagen, Denmark}
\acmBooktitle{Proceedings of the 21st Workshop on Privacy in the Electronic Society (WPES '23) (to appear), November 26, 2023, Copenhagen, Denmark}
\acmPrice{}
\acmISBN{979-8-4007-0235-8/23/11}

\RequirePackage[
  datamodel=acmdatamodel,
  style=acmnumeric,
  ]{biblatex}

\newcommand{\qm}[1]{``#1''}

\newtheorem{theorem}{Theorem}

\newtheorem{lemma}[theorem]{Lemma}

\newtheorem{remark}[theorem]{Remark}

\begin{document}

\title{A Quantitative Information Flow Analysis of the Topics API}

\author{Mário S. Alvim}
\email{msalvim@dcc.ufmg.br}
\orcid{0000-0002-4196-7467}
\affiliation{%
  \institution{Universidade Federal de Minas Gerais}
  \city{Belo Horizonte}
  \state{MG}
  \country{Brazil}
}

\author{Natasha Fernandes}
\email{natasha.fernandes@mq.edu.au}
\orcid{0000-0002-9212-7839}
\author{Annabelle McIver}
\email{annabelle.mciver@mq.edu.au}
\orcid{0000-0002-2405-9838}
\affiliation{%
  \institution{Macquarie University}
  \city{Sydney}
  \state{NSW}
  \country{Australia}}

\author{Gabriel H. Nunes}
\email{ghn@nunesgh.com}
\orcid{0000-0002-7823-3061}
\affiliation{%
  \institution{Macquarie University}
  \city{Sydney}
  \state{NSW}
  \country{Australia}
}
\affiliation{%
  \institution{Universidade Federal de Minas Gerais}
  \city{Belo Horizonte}
  \state{MG}
  \country{Brazil}
}

\renewcommand{\shortauthors}{Alvim et al.}

\begin{abstract}
    Third-party cookies have been a privacy concern since cookies were first developed in the mid 1990s, but more strict cookie policies were only introduced by Internet browser vendors in the early 2010s.
    More recently, due to regulatory changes, browser vendors have started to completely block third-party cookies, with both Firefox and Safari already compliant.

    The \emph{Topics API} is being proposed by Google as an additional and less intrusive source of information for \emph{interest-based advertising} (IBA), following the upcoming deprecation of third-party cookies.
    Initial results published by Google estimate the probability of a correct re-identification of a random individual would be below 3\% while still supporting IBA.

    In this paper, we analyze the re-identification risk for individual Internet users introduced by the Topics API from the perspective of \emph{Quantitative Information Flow} (QIF), an information- and decision-theoretic framework.
    Our model allows a theoretical analysis of both privacy and utility aspects of the API and their trade-off, and we show that the Topics API does have better privacy than third-party cookies.
    We leave the utility analyses for future work.
\end{abstract}







\maketitle

\section{Introduction}
\label{sec:introduction}

\paragraph{Third-party cookies.}
Cookies were first formally specified by the \emph{Internet Engineering Task Force} (IETF) in 1997 as \qm{a way to create a stateful session with HTTP requests and responses} \cite{Montulli1997}.
From their inception, it was known that cookies were vulnerable to privacy abuse through \qm{cookie sharing}, now known as \emph{third-party cookies}, and Internet browser vendors were strongly encouraged to \qm{prevent the sharing of session information between hosts that are in different domains} \cite{Montulli1997}, but more strict cookie policies were only introduced by Internet browser vendors in the early 2010s \cite{Mayer2013}.

\begin{figure*}[t]
    \centering
    \begin{subfigure}[b]{0.45\textwidth}
        \includegraphics[width=\textwidth]{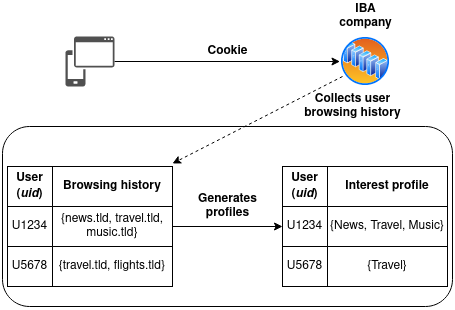}
        \caption{An IBA company using third-party cookies.}
        \label{fig:cookies}
    \end{subfigure}
    \hfil
    \begin{subfigure}[b]{0.45\textwidth}
        \includegraphics[width=\textwidth]{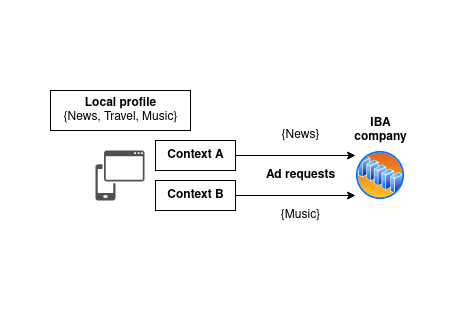}
        \caption{An IBA company using the Topics API.}
        \label{fig:topics}
    \end{subfigure}
    \caption{Information collected by an interest-based advertising (IBA) company.}
    \label{fig:schematics}
\end{figure*}

A \emph{cookie} is a piece of information stored by an Internet browser that consists of a tuple of the \emph{origin} (domain) that set that cookie and one or more pairs of keys and values.
For instance, a cookie may be used to store the content of a shopping cart on a e-commerce origin, or the login information on a social network or e-mail service.
But a cookie may also be set by any third-party origin called on a \emph{context} (visited web page), e.g. through advertisements or social widgets.
If a third-party sets a cookie with a \emph{uid} (unique identifier) on a browser, it becomes capable of tracking the browsing history of that individual on that browser whenever the third-party is called across the Internet, as depicted in Figure~\ref{fig:cookies}, which enables the creation of precise browsing profiles for individuals at scale.
Due to regulatory changes, browser vendors have finally started to deprecate third-party cookies \cite{Wilander2020,Wood2019}.

\paragraph{Topics API}
Given the deprecation of third-party cookies, the Topics API is being proposed by Google to provide third-parties with \qm{coarse-grained advertising topics that the page visitor might currently be interested in} \cite{Google2022}.
This includes \emph{interest-based advertising} (IBA), which differs from contextual advertising by taking into account not only the context for deriving an individual's set of interests, but also additional signals, such as those provided by browsing profiles created with third-party cookies.

The Topics API proposes the representation of an individual as a set of top interests derived from their browsing history, a pre-trained classification model, and a pre-defined taxonomy of interests.
Roughly, at the end of each week, the individual's browser would locally compute a fixed-size set of top topics of interest based on the browsing history and the topics assigned to each of the observed contexts.
Once computed, this set would be available to third-parties for a fixed number of weeks and under certain restrictions, as depicted in Figure~\ref{fig:topics}.

For instance, a third-party would receive only one topic per individual, per week, and per context, chosen uniformly at random from that individual's set of top topics for that week.
Moreover, a third-party would not receive a topic unrelated to the contexts it has witnessed that individual visit on that week, and, with a 5\% chance, the received topic would be instead chosen uniformly at random from the whole taxonomy \cite{Carey2023,Google2022}.

\paragraph{Quantitative Information Flow.}
QIF, an information- and decision-theoretic framework \cite{Alvim2020}, models secrets as probability distributions, $\pi$, that represent an adversary's knowledge on which value the secret is more likely to take.
QIF is flexible when it comes to choosing vulnerability or entropy measures, i.e. how much damage is caused by the leakage of information or how valuable that information is to an adversary, respectively.

QIF models systems that process information as information-theoretical channels, usually represented as stochastic channel matrices $\mathsf{C}$ that map from a (finite) set of secret inputs $\mathcal{X}$ to a (finite) set of observable outputs $\mathcal{Y}$.
Hence, each entry $\mathsf{C}_{x,y}$ denotes $p(y \mid x)$, the conditional probability of getting output $y$ given input $x$.
It is important to note that QIF makes the worst-case assumption that the adversary knows how the channel works, i.e. the adversary knows the channel matrix $\mathsf{C}$, or at least is capable of computing $\mathsf{C}_{x,y}$ for any $x \in \mathcal{X}$, $y \in \mathcal{Y}$.

Given a channel matrix representation of a system and an initial probability distribution on the secret, QIF describes the effect of the channel on the adversary's knowledge about the secret as a hyper-distribution (\emph{hyper}).
A hyper, $[\pi \triangleright \mathsf{C}]$, is a probability distribution (\emph{outer distribution}) on final probability distributions (\emph{inner distributions}), each representing a possible, updated state of knowledge the adversary might have after observing the outputs from the system.

QIF then considers all the possible outputs from the channel independently from any particular execution of the system and defines the final vulnerability as the expected value of the appropriate vulnerability measure over the hyper, i.e. the vulnerabilities of the inner distributions weighted by their respective outer probabilities.~\footnote{In QIF, the information-theoretic essence of a channel matrix $\mathsf{C}$ is a mapping from initial probability distributions $\pi$ to hypers $[\pi \triangleright \mathsf{C}]$ \cite{Alvim2020}.}
Finally, the leakage of information can be computed in absolute (\emph{additive}) or relative (\emph{multiplicative}) terms as a comparison between the initial vulnerability, i.e. the vulnerability of only the adversary's initial probability distribution on the secret, and the final vulnerability.

QIF has already been successfully applied to a variety of privacy and security analyses, including searchable encryption \cite{Jurado2019}, intersection and linkage attacks against \emph{k}-anonymity \cite{Fernandes2018}, privacy analysis of very large datasets \cite{Alvim2022}, and differential privacy \cite{Alvim2015,Chatzikokolakis2019}.





\paragraph{Objectives.}
From an individual Internet user's perspective, the existence of third-party cookies constitutes a serious privacy risk that is available to any third-party capable of setting them.
But from IBA companies' perspective, third-party cookies have been a fine-grained source of information for deriving individuals' interests for targeted advertising.
The Topics API is expected to \qm{support IBA without relying on cross-site tracking} \cite{Carey2023} by making it harder for third-parties to link topics observed on different contexts to the same individual, if compared to the direct linkage via \emph{uid}s enabled by third-party cookies.
In fact, initial results published by Google estimate the probability of a correct re-identification of a random individual would be below 3\% \cite{Carey2023}.

Our goal is to formally verify the claims made by the proponents of the Topics API by developing a sound yet easily explainable model for both the API and third-party cookies, which we use as a baseline for comparison.
We are interested in understanding both the privacy (for Internet users) and utility (for IBA companies) implications of the migration from third-party cookies to the Topics API, how they relate, and their trade-off.
In this paper, we analyze the re-identification risk for individual Internet users.
We leave the utility analysis for future work.

\paragraph{Contributions.}
We provide a novel and rigorous QIF model that allows a theoretical analysis of the privacy aspects of the Topics API.
Moreover, our model is flexible enough to allow additional adversarial scenarios not considered in this paper, such as attribute-inference or longitudinal attacks \cite{Alvim2022}, and to allow the analysis of the utility aspects and of the privacy-utility trade-off of the API \cite{Alvim2020a}.



\section{Analysis}
\label{sec:analysis}

\begin{table*}[t]
    \centering
    \begin{subtable}[b]{0.45\textwidth}
        \centering
        \begin{tabular}{c|c c c c}
            $\mathsf{C}_{\mathcal{C}}$ & $\{ \mathcal{D}_1, \mathcal{D}_2 \}$ & $\{ \mathcal{D}_1, \mathcal{D}_3 \}$ & $\{ \mathcal{D}_2, \mathcal{D}_3 \}$ & $\{ \mathcal{D}_1, \mathcal{D}_2, \mathcal{D}_3 \}$ \\
            \hline
            $x_1$ & 0 & 1 & 0 & 0 \\
            $x_2$ & 1 & 0 & 0 & 0 \\
            $\vdots$ & $\vdots$ & $\vdots$ & $\vdots$ & $\vdots$ \\
            $x_{N - 1}$ & 0 & 1 & 0 & 0 \\
            $x_N$ & 0 & 0 & 0 & 1
        \end{tabular}
        \caption{Channel for the third-party cookies model considering only three possible contexts $\mathcal{D}_i$. Each row represents a distinct Internet user $x_i$ and the corresponding browsing history consisting of a combination of the possible contexts $\mathcal{D}_i$. As a stochastic channel, every and each row sums to 1. For instance, user $x_1$ has a browsing history consisting of the contexts $\mathcal{D}_1$ and $\mathcal{D}_3$.}
        \label{tab:cookies-channel}
    \end{subtable}
    \hfil
    \begin{subtable}[b]{0.45\textwidth}
        \centering
        \begin{tabular}{c|c c c c c c c}
            $\mathsf{C}_{\mathcal{T}}$ & $t_1$ & $t_2$ & $t_3$ & $\cdots$ & $t_{j - 2}$ & $t_{j - 1}$ & $t_j$ \\
            \hline
            $x_1$ & $\nicefrac{1}{k}$ & 0 & $\nicefrac{1}{k}$ & $\cdots$ & $\nicefrac{1}{k}$ & $\nicefrac{1}{k}$ & 0 \\
            $x_2$ & $\nicefrac{1}{k}$ & $\nicefrac{1}{k}$ & 0 & $\cdots$ & 0 & $\nicefrac{1}{k}$ & $\nicefrac{1}{k}$ \\
            $\vdots$ & $\vdots$ & $\vdots$ & $\vdots$ & $\vdots$ & $\vdots$ & $\vdots$ & $\vdots$ \\
            $x_{N - 1}$ & $\nicefrac{1}{k}$ & $\nicefrac{1}{k}$ & $\nicefrac{1}{k}$ & $\cdots$ & 0 & 0 & 0 \\
            $x_N$ & $\nicefrac{1}{k}$ & 0 & $\nicefrac{1}{k}$ & $\cdots$ & 0 & 0 & $\nicefrac{1}{k}$
        \end{tabular}
        \caption{Channel for the Topics API model. Each row represents a distinct Internet user $x_i$ and the corresponding probability for each of the $t_i$ topics on the columns. As a stochastic channel, every and each row sums to 1. For instance, we know that user $x_1$ has at least topics $t_1$, $t_3$, $t_{j-2}$, and $t_{j-1}$ in their set of \emph{k}-top topics of interest.}
        \label{tab:topics-channel}
    \end{subtable}
    \caption{Channels for the third-party cookies and the Topics API models.}
    \label{tab:channels}
\end{table*}

We start by defining our adversary and the initial vulnerability for an Internet user.
Then, in Section~\ref{sec:cookies}, we model third-party cookies, which we use as a baseline for comparison.
Next, in Section~\ref{sec:topics}, we model the Topics API.
Finally, in Section~\ref{sec:comparison}, we compare the re-identification risks for an Internet user under each scenario.


\paragraph{Initial vulnerability.}
We consider the adversary is interested in learning the identity of Internet users, i.e. re-identifying the users, each of whom has just visited two distinct websites on a newly installed Internet browser.
Hence, our secret is the identity of Internet users and each user is an element $x$ of the finite set $\mathcal{X}$.

Lacking any tracking or fingerprinting capabilities, the adversary cannot distinguish between any individual and all other individuals on the Internet.
Hence, the adversary's initial probability distribution on individuals' identities is modeled as a uniform distribution, $\pi = \vartheta = \nicefrac{1}{N}$, where $N = |\mathcal{X}|$ is the total number of Internet users.

We are interested in measuring how likely an adversary is to correctly re-identify an Internet user at first try.~\footnote{A lower bound for multiple tries.}
Such scenario can be modeled by the Bayes vulnerability measure, defined as $V_1 (\pi) \coloneqq \max_{x \in \mathcal{X}} \pi_x$ \cite[Def. 2.3]{Alvim2020}, where $\pi$ is a probability distribution on a finite set $\mathcal{X}$.
Moreover, we assume an \emph{information theoretic} adversary, i.e. without bounds on computational resources.




\begin{remark}[Initial vulnerability]
    \label{rem:prior-vulnerability}
    Given the adversary's uniform initial probability distribution and the Bayes vulnerability measure, the initial vulnerability is:
    \begin{align}
        V_1 (\pi) = V_1 (\vartheta) = \nicefrac{1}{N},
    \end{align}
    where $N = |\mathcal{X}|$ is the total number of Internet users.
\end{remark}

\paragraph{Final vulnerability.}
In order to compute the final Bayes vulnerability, i.e. on the hyper distribution instead of the initial probability distribution, we rely on two theorems.
For a deterministic channel and a uniform initial probability distribution, it is known that the final Bayes vulnerability equals $V_1 [\vartheta \triangleright \mathsf{C}] = \nicefrac{M}{N}$, where $\vartheta$ is a uniform probability distribution on an $N$-element set $\mathcal{X}$, and $\mathsf{C}$ is a deterministic channel with $M$ possible output values \cite[Theo. 1.1]{Alvim2020}.
For any channel and initial probability distribution, it is known that the final Bayes vulnerability equals $V_1 [\pi \triangleright \mathsf{C}] = \sum_{y \in \mathcal{Y}} \max_{x \in \mathcal{X}} \mathsf{J}_{x,y}$, where $\mathsf{J} = \pi \triangleright \mathsf{C}$ is the joint matrix, i.e. the final Bayes vulnerability equals the sum of the column maximums of the joint matrix $\mathsf{J}$ \cite[Theo. 5.15]{Alvim2020}.





\subsection{Third-party cookies vulnerability}
\label{sec:cookies}

We assume a powerful adversary capable of reconstructing the whole browsing histories of every Internet user.
Therefore, the channel is a mapping of Internet users $x_i$ to browsing histories, each containing a combination of at least two contexts and at most every possible context, i.e. a deterministic channel as depicted on Table~\ref{tab:cookies-channel} considering only three possible contexts $\mathcal{D}_i$ (for domain).


Such a powerful adversary is justified based on previous findings reported on the literature.
For instance, nearly 90\% of the 500 most popular websites in 2011 included at least one third-party known for tracking users' browsing histories, with the most common third-party at the time present on almost 40\% of them \cite{Roesner2012}.
Moreover, when considering the possibility of collusion among third-parties, the top 10 advertising and analytics companies in 2018 could observe more than 90\% of users' browsing history \cite{Bashir2018}.



\begin{lemma}[Final vulnerability for third-party cookies]
    \label{lem:cookies-posterior-vulnerability}
    Given the adversary's uniform initial probability distribution on individuals, the deterministic channel mapping Internet users to browsing histories, and the Bayes vulnerability measure, the final vulnerability for third-party cookies is:
    \begin{align}
        V_1 [\vartheta \triangleright \mathsf{C}_{\mathcal{C}}] = \frac{1}{N} \sum_{k'=2}^{k'=M'} \binom{M'}{k'} = \frac{1}{N} \sum_{k'=2}^{k'=M'} \frac{M'!}{k'! (M' - k')!},
    \end{align}
    where $M'$ is the number of contexts on the Internet that include third-party cookies, $k'$ is the number of contexts on an Internet user's browsing history that may be affected by third-party cookies linkage, and $N = |\mathcal{X}|$ is the total number of Internet users.
\end{lemma}



We will use the result from Lemma~\ref{lem:cookies-posterior-vulnerability} as a baseline for comparison with the proposed Topics API.
It is important to note that, as the number of contexts increase, the channel from Table~\ref{tab:cookies-channel} assumes the shape of the identity matrix, known in the theory of QIF as the channel that \emph{annihilates} secrecy, i.e. that leaks everything.

\subsection{Topics API vulnerability}
\label{sec:topics}

We assume a powerful adversary capable of reconstructing the whole set of \emph{k}-top topics for every Internet user in a given week, where $k \geq 1$ is an integer.
We do not consider, for now, the 5\% chance of the received topic being instead chosen uniformly at random from the whole taxonomy.~\footnote{This could be easily included in our model through channel composition, but we leave this for future work.}
Therefore, the channel is a mapping of Internet users $x_i$ to topics $t_i$, and each entry of the matrix is either $\nicefrac{1}{k}$, if the corresponding user has the corresponding topic in their set of \emph{k}-top topics, or $0$ otherwise, as depicted on Table~\ref{tab:topics-channel}.

Such a powerful adversary is justified based on the possibility of collusion among callers of the Topics API in order to link the identity of Internet users across them \cite{Carey2023}.

This channel represents the use of two anonymization techniques known from the literature, i.e. \emph{k}-anonymity's generalization \cite{Sweeney2000} on the classification of contexts according to topics, and bounded noise on the limitation on the size of the set of \emph{k}-top topics.
Even though both techniques have been widely used for statistical data publications, they have already been shown to be vulnerable to re-identification attacks, such as linkage \cite{Machanavajjhala2007} and composition \cite{Ganta2008} attacks against \emph{k}-anonymity and histogram reconstruction \cite{Asghar2020} against bounded noise.

\begin{lemma}[Final vulnerability for the Topics API]
    \label{lem:topics-posterior-vulnerability}
    Given the adversary's uniform initial probability distribution on individuals, the channel mapping Internet users to topics, and the Bayes vulnerability measure, the final vulnerability for the Topics API is:
    \begin{align}
        V_1 [\vartheta \triangleright \mathsf{C}_{\mathcal{T}}] = \frac{1}{N} \frac{M}{k},
    \end{align}
    where $N = |\mathcal{X}|$ is the total number of Internet users, $k$ is the size of the \emph{k}-top topics set, and $M$ is the size of the finite set of topics that have at least one non-zero occurrence, i.e. $M$ is the total number of topics observed for all Internet users in the considered week.~\footnote{In fact, the total number of topics \emph{observed} for all Internet users in the considered week equals the total number of topics \emph{implemented} by the API. The current draft proposal of the Topics API introduces \qm{a 5\% chance that a per-user, per-site, per-epoch random topic is returned (chosen uniformly at random)} \cite{Google2022} when the API is called. \qm{The 5\% noise is introduced to ensure that each topic has a minimum number of members (\emph{k}-anonymity) as well as to provide some amount of plausible deniability} \cite{Google2022}. Therefore, $j = M$ on Table~\ref{tab:topics-channel}.}
\end{lemma}


\subsection{Comparing leakages}
\label{sec:comparison}

According to the current draft proposal for the Topics API, the taxonomy would consist of about 350 topics, the size of the \emph{k}-top topics set would be equal to five, and the number of weeks accessible through the API would be equal to three \cite{Google2022}.
It is important to note that those values are subject to change over time, as well as the classification model and the taxonomy of interests, which would directly impact the vulnerability of the Topics API.

We consider here the QIF notion of \emph{multiplicative leakage}, i.e. a relative comparison between the initial and final vulnerabilities defined as $\mathcal{L}^{\times}_1 (\pi, C) \coloneqq \nicefrac{V_1 [\pi \triangleright C]}{V_1 (\pi)}$ for Bayes vulnerability.

For the Topics API, considering its current default parameters and the results from Remark~\ref{rem:prior-vulnerability} and Lemma~\ref{lem:topics-posterior-vulnerability} for the initial and final vulnerabilities, respectively, we would have a leakage equal to
\begin{align}
    \mathcal{L}^{\times}_1 (\vartheta, \mathsf{C}_{\mathcal{T}}) = \frac{M}{k} = \frac{350}{5} = 70,
\end{align}
i.e. the adversary would have their knowledge increased by 70 times after using the Topics API, when starting from a uniform probability distribution.

\begin{figure}[t]
    \centering
    \includegraphics[width=0.45\textwidth]{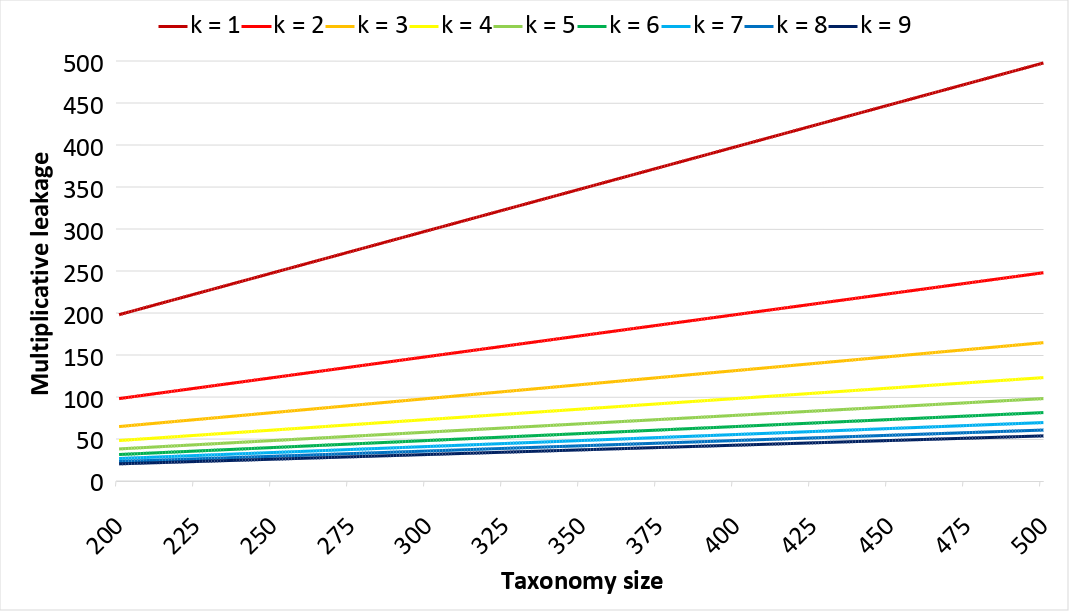}
    \caption{Topics API multiplicative leakage given the size of the taxonomy for different sizes of the \emph{k}-top topics list.}
    \label{fig:topics-leakage}
\end{figure}

Figure~\ref{fig:topics-leakage} depicts the multiplicative leakage of the Topics API given the size of the taxonomy, $M$, and for different sizes of the \emph{k}-top topics list, $k$, according to the results from Remark~\ref{rem:prior-vulnerability} and Lemma~\ref{lem:topics-posterior-vulnerability}.
As expected, information leakage \emph{increases} as more topics are included in the taxonomy or as less topics are required for each users' \emph{k}-top topics list.

For third-party cookies, considering only the 500 most visited websites on the Internet and the results from Remark~\ref{rem:prior-vulnerability} and Lemma~\ref{lem:cookies-posterior-vulnerability} for the initial and final vulnerabilities, respectively, we would have a leakage equal to $\mathcal{L}^{\times}_1 (\vartheta, \mathsf{C}_{\mathcal{C}}) = \sum_{k'=2}^{k'=M'} \nicefrac{M'!}{k'! (M' - k')!} = 1.8 \cdot 10^{238}$.
But not all websites are used for third-party tracking, so considering that the most common third-party back in 2011 was present on almost 40\% of the 500 most visited websites at the time \cite{Roesner2012}, we would have a leakage equal to
\begin{align}
    \mathcal{L}^{\times}_1 (\vartheta, \mathsf{C}_{\mathcal{C}}) = \sum_{k'=2}^{k'=M'} \frac{M'!}{k'! (M' - k')!} = 1.3 \cdot 10^{95},
\end{align}
i.e. the adversary would have their knowledge increased by $10^{95}$, when starting from a uniform probability distribution.

This result makes it blatantly clear that third-party cookies have the potential to reveal everything and can indeed be used for the re-identification of individuals at scale, in agreement with Sweeney's results from the early 2000s that showed that individuals are re-identifiable with only three pieces of information \cite{Sweeney2000}.






\section{Conclusion}
\label{sec:conclusion}
In this paper, we have provided a novel and rigorous QIF model that allows a theoretical analysis of the privacy aspects of the Topics API, including how much the API's leakage changes as we vary the API's default parameters, i.e. the sizes of the taxonomy and of the \emph{k}-top topics list, and a comparative analysis of the API's leakage with third-party cookies as a baseline.
Moreover, our model is flexible enough to allow additional adversarial scenarios not considered in this paper \cite{Alvim2022}, and to allow the analysis of utility aspects and of the privacy-utility trade-off of the Topics API \cite{Alvim2020a}.

We have shown that the proposed Topics API is a clear improvement over third-party cookies, particularly under a very powerful adversary.
Nevertheless, re-identification of individuals is still possible \cite{Carey2023} and more formal and experimental results are needed to fully assess the vulnerabilities intrinsic to the Topics API, particularly considering the techniques used in its conception, which are known to be vulnerable against re-identification attacks \cite{Machanavajjhala2007,Ganta2008,Asghar2020}.



\section{Acknowledgments}
Mário S. Alvim and Gabriel H. Nunes were partially funded by CNPq, CAPES, and FAPEMIG.

\balance
\printbibliography

\end{document}